\documentclass[prl,twocolumn,showpacs,showkeys]{revtex4}

\usepackage{amssymb, amsmath, amsfonts, latexsym, verbatim, mathrsfs}

\usepackage{graphicx}

\begin{document}

\title{Resonant purification of mixed states for closed and open quantum systems}
\thanks{This material is based upon work supported by
the National Science Foundation under Grant No. 0237925}

\author{Raffaele Romano}
\email{rromano@iastate.edu} \affiliation{Department of Mathematics,
Iowa State University, Ames IA 50011, USA}


\begin{abstract}
\noindent Pure states are fundamental for the implementation of
quantum technologies, and several methods for the purification of
the state of a quantum system $S$ have been developed in the past
years. In this letter we present a new approach, based on the
interaction of $S$ with an auxiliary system $P$, having a wide range
of applicability. Considering two-level systems $S$ and $P$ and
assuming a particular interaction between them, we prove that
complete purifications can be obtained under suitable conditions on
the parameters characterizing $P$. Using analytical and numerical
tools, we show that the purification process exhibits a resonant
behavior in both the cases of system isolated from the external
environment or not.

\end{abstract}



\maketitle


{\it Introduction.---} The quantum theories of information and
computation suggest that quantum mechanics based devices could
highly outperform the corresponding classical apparatuses in several
fields~\cite{niel}. At the heart of this qualitative improvement
there are the quantum correlations called {\it entanglement}, absent
in classical systems. In full generality, {\it pure states} are
needed for the implementation of quantum processing. These states,
represented by projectors on the Hilbert space associated to the
system, do not contain classical correlations.

Unfortunately, when a quantum system interacts with its surrounding
environment, it loses its relevant properties, since it is subject
to a dissipative evolution leading to decoherence and irreversible
transitions from pure to {\it mixed states} (that is probabilistic
superpositions of pure states). To avoid this problem, it is
fundamental to decouple as much as possible system and environment
or, whenever this is not possible, to study methods for
counteracting the environmental action, thus preserving the
entanglement and restoring the purity of the states (the so-called
{\it purification} process).

The study of methods for the purification of a quantum state, that
is the transition to less mixed states, provides a rich field of
investigation. Several strategies have already been proposed,
especially in the context of entanglement
purification~\cite{benn,cira}. They rely on two mechanisms:
measurement or cooling procedures
(see~\cite{naka,comb,wise2,tann,skla} and references therein). From
a general viewpoint, it is necessary to consider how it is possible
to affect the dynamics of the system, and what states can be
attained during its time evolution. In this context, one assumes
that some dynamical parameters can be externally modified. They are
called {\it controls} and, by means of them, specific transitions
can be induced (see~\cite{rama,albe} and references therein).

Usually, the controls are assumed to affect the dynamics through the
Hamiltonian of the system. The corresponding models are called
coherent control methods as the controls enter the coherent part of
the dynamics, and their properties have been widely investigated. In
particular, it has been proven that quantum states cannot be
purified by using coherent control, both in closed and open systems
dynamics, whether the controls are fixed at the
beginning~\cite{kett,alta}. To overcome this difficulty, several
solutions have been proposed, as the use of an indirect measurement,
performed on the system. In some schemes, the outcomes of the
measurement are used to continuously update the controls ({\it
quantum feedback}, see e.g. \cite{wise,manc}), and state
purifications can be realized. Another possibility is represented by
the cooling techniques: they rely on a constructive interplay of
coherent control and specific dissipative mechanisms, leading to a
purity increase in the system at the bath expenses.

Recently, a different control scheme has been developed, in which
the controls are fixed {\it a priori}, and the assumption of
coherent control is relaxed~\cite{roma}. This non-coherent control
protocol relies on the use of an initially uncorrelated auxiliary
system, the probe $P$, that is put in interaction with the relevant
system $S$ and discarded at the end of the procedure. The controls
$u$ enter the dynamics through the initial state of the probe,
$\rho_P (0) = \rho_P (u)$, and the ability to modify the states of
the target system depends on the correlations between system and
probe, generated by the interaction. The dynamics of the system is
given by
\begin{equation}\label{eq01}
    \rho_S (t, u) = {\rm Tr}_P \rho_T (t)
\end{equation}
with
\begin{equation}\label{eq011}
    \rho_T (t) = \gamma_t [\rho_S (0) \otimes \rho_P (u)],
\end{equation}
where $T$ denotes the composite system $S + P$, and $\gamma_t$ its
time evolution. If $T$ is a closed system, $\gamma_t$ is generated
by the Liouville operator
\begin{equation}\label{eq012}
    \dot{\rho}_T (t) = -i [H_T, \rho_T (t)],
\end{equation}
where $H_T = H_S + H_P + H_I$ is the total Hamiltonian, containing
the free contributions $H_S$ and $H_P$ and the interaction term
$H_I$. In principle, there is some freedom in the choice of the
operators $H_P$ and $H_I$, because different probe systems could be
considered.

In this letter we prove that the non-coherent control protocol
provides an alternative approach to the purification of a qubit
system. For a particular choice of the free dynamics of the probe
and its interaction with the system, we prove that this control
scheme leads to complete purifications of the maximally mixed state
also in the pessimistic case of weak interaction between $S$ and
$P$, if there is not a surrounding environment. By using both
analytical and numerical tools, we find that the purification
process has a resonant behavior depending on the energy difference
between the two systems. Finally, by considering the worst-case
isotropic decoherence, we show how our results are affected by the
action of the external environment.

We limit our attention to two-dimensional $S$ and $P$. As a measure
of the purity of a state $\rho_S$ we consider the von Neumann
distance of this state from the maximally mixed state ${\mathbb
I}/2$,
\begin{equation}\label{eq01tris}
    \pi = \sqrt{2 {\rm Tr} \Bigl(\rho_S -
    \frac{\mathbb I}{2}\Bigr)^2} = \sqrt{2 {\rm Tr} (\rho_S^2) - 1}
\end{equation}
where a convenient scale factor $2$ has been included, such that $0
\leqslant \pi \leqslant 1$, $\pi = 0$ if and only if $\rho_S$ is the
maximally mixed state, and $\pi = 1$ if and only if $\rho_S$ is a
pure state, $\rho_S^2 = \rho_S$.

We introduce a Bloch vector representation for the states,
\begin{equation}\label{eq02}
    \rho_S (t) = \frac{1}{2} \bigl( {\mathbb I} + \vec{s} (t) \cdot
    \vec{\sigma}^S \bigr), \quad \rho_P (t) = \frac{1}{2} \bigl(
    {\mathbb I} + \vec{p} (t) \cdot \vec{\sigma}^P \bigr)
\end{equation}
with $\vec{s} (t)$, $\vec{p} (t)$ real vectors satisfying $\|
\vec{s} (t) \| = \| \vec{p} (t) \| \leqslant 1$ and
$\vec{\sigma}^S$, $\vec{\sigma}^P$ the vectors of Pauli matrices in
$S$ and $P$ respectively. The time-dependent purity becomes
\begin{equation}\label{eq03}
    \pi(t) = \| \vec{s} (t) \|,
\end{equation}
and a purification process amounts to an increase of $\pi (t)$.

For coherent control, $\dot{\rho}_S (t) = -i [H_S (u), \rho_S (t)]$,
therefore
\begin{equation}\label{eq03bis}
    \dot{\pi} (t) = 2 \pi^{-1} (t) {\rm Tr} \bigl(\rho_S (t)
\dot{\rho}_S (t)\bigr) = 0,
\end{equation}
and the purity of the initial state is not modified, as it has been
remarked before.


{\it Purification by means of a probe.---} If $T$ is a closed
system, the operator $\gamma_t$ in (\ref{eq011}) is given by
\begin{equation}\label{eq01bis}
    \gamma_t [\,\cdot\,] = e^{-i H_T t} \,\cdot\, e^{i H_T t}.
\end{equation}
In general, it is possible to explore the purification process by
using a Cartan decomposition,
\begin{equation}\label{eq03tris}
    e^{-i H_T t} = L_1 e^{a t} L_2
\end{equation}
where $L_1$, $L_2$ are local contributions and $a = c_x \sigma_x^S
\otimes \sigma_x ^P + c_y \sigma_y^S \otimes \sigma_y ^P + c_z
\sigma_z^S \otimes \sigma_z ^P$, with $c_x, c_y, c_z \in {\mathbb
R}$. Since the local transformations do not affect the purification
process, it is possible to classify all the possible cases by means
of $c_x, c_y$ and $c_z$. In fact, the entangling capability of the
operator (\ref{eq03tris}), at the heart of the non-coherent control
model, depends on these coefficients. By using the expression of
$\vec{s} (t)$ derived in \cite{roma}, we find that evolutions
leading to an increase of the purity of the initial state are
possible only if at least two different coefficients $c_i$ do not
vanish. This condition is weaker than the accessibility and
controllability criteria, thus systems that are neither accessible
nor controllable can exhibit purification of mixed states. Moreover,
it is possible to prove that whenever the maximally mixed state can
be completely purified, the same is true for an arbitrary mixed
state.

Our aim in this work is to discuss some relevant features of the
purification process under the incoherent control protocol, rather
than to systematically study all the possible cases. Moreover, we
want to analyze this process in terms of the relevant physical
quantities entering the dynamics (the characteristic energies of $S$
and $P$), rather than in terms of the $c_i$ parameters. Therefore we
consider a particular model by assuming $H_S = \omega_s \sigma_z^S$,
$H_P = \omega_p \sigma_z^P$ and $H_I = g \sigma_x^S \otimes
\sigma_x^P$, where $\omega_s$ and $\omega_p$ are the characteristic
energies in $S$ and $P$, and $g$ is the coupling constant. The
propagator for the total system can be directly computed as
\begin{eqnarray}\label{eq04}
    e^{-i H_T t} &=& \frac{1}{2} \cos{(\alpha t)} ({\mathbb I} \otimes
    {\mathbb I} + \sigma_z^S \otimes \sigma_z^P) + \nonumber \\
    &+& \frac{1}{2} \cos{(\beta t)} ({\mathbb I} \otimes
    {\mathbb I} - \sigma_z^S \otimes \sigma_z^P) + \nonumber \\
    &-& i \frac{a}{2 \alpha} \sin{(\alpha t)} (\sigma_z^S \otimes
    {\mathbb I} + {\mathbb I} \otimes \sigma_z^P) + \nonumber \\
    &-& i \frac{b}{2 \beta} \sin{(\beta t)} (\sigma_z^S \otimes
    {\mathbb I} - {\mathbb I} \otimes \sigma_z^P) + \nonumber \\
    &-& i \frac{g}{2 \alpha} \sin{(\alpha t)} (\sigma_x^S \otimes
    \sigma_x^P - \sigma_y^S \otimes \sigma_y^P) + \nonumber \\
    &-& i \frac{g}{2 \beta} \sin{(\beta t)} (\sigma_x^S \otimes
    \sigma_x^P + \sigma_y^S \otimes \sigma_y^P),
\end{eqnarray}
where
\begin{eqnarray}\label{eq05}
  \bar{\omega} &=& \omega_s + \omega_p, \qquad \alpha = \sqrt{\bar{\omega}^2 + g^2}, \nonumber \\
  \delta \omega &=& \omega_s - \omega_p, \qquad \beta = \sqrt{(\delta \omega)^2 + g^2}.
\end{eqnarray}

Following the previous discussion, it is not restrictive to choose
the maximally mixed state $\rho_S (0) = {\mathbb I}/2$ as initial
state for the system $S$. The initial state of the composite system
is given by $\rho_T (0) = \rho_S (0) \otimes \rho_P (u)$. Denoting
by $\vec{s} (0) = (0, 0, 0)$ and $\vec{p} (u) = (p_x, p_y, p_z)$ the
Bloch vector representations of the initial states $\rho_S (0)$ and
$\rho_P (u)$, we can compute $s_i = {\rm Tr} \bigl( \rho_T (t)
\sigma_i^S \otimes {\mathbb I} \bigr)$ for $i = x, y, z$, with
$\rho_T (t)$ from (\ref{eq01bis}),
\begin{eqnarray}\label{eq06}
  s_x (t) &=& s_y (t) = 0, \nonumber \\
  s_z (t) &=& p_z g^2 \Bigl( \frac{1}{\beta^2} \sin^2{(\beta t)} -
  \frac{1}{\alpha^2} \sin^2{(\alpha t)} \Bigr).
\end{eqnarray}

\begin{figure}[t]
\begin{center} 
  \includegraphics[width=8cm]{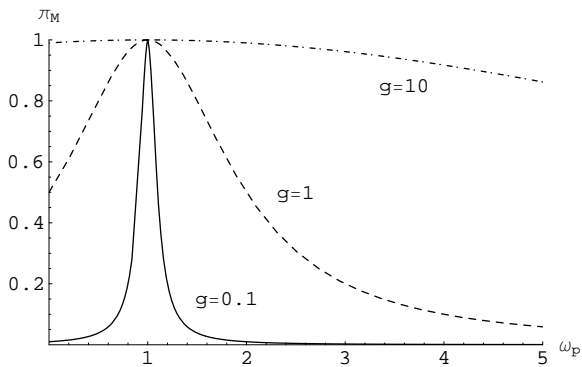} \\
 \caption{\footnotesize Maximal attainable purification $\pi_M$ for the maximally
 mixed state in several regimes. The plots exhibit a resonance about
 $\omega_p = \omega_s$ in all the cases: dominant interaction (dot-dashed line),
 interaction and free contribution of the same strength (dashed line), weaker interaction
 (solid line). The value $\omega_s = 1$ has been considered.}\label{fig1}
\end{center}
\end{figure}

As the $x$ and $y$ components are constant, the system is not
accessible nor controllable. The purity of the state is given by
$\pi (t) = \vert s_z (t) \vert$ and it is a combination of
oscillating functions. After some manipulations, it can be written
in the form
\begin{eqnarray}\label{eq06bis}
    s_z (t) &=& p_z g^2 \Bigl[ \frac{\alpha^2 -
    \beta^2}{2 \alpha^2 \beta^2} \Bigl( 1 - \cos{(\alpha + \beta)t}
    \cos{(\alpha - \beta)t} \Bigr) + \nonumber \\
    &-& \frac{\alpha^2 + \beta^2}{2 \alpha^2 \beta^2} \sin{(\alpha +
    \beta)t} \sin{(\alpha - \beta)t} \Bigr].
\end{eqnarray}
We are interested in the maximal amplitude of this function, that we
denote by $\pi_M$. Unfortunately, it is not possible to give an
analytical expression of $\pi_M$, unless the interaction between $S$
and $P$ is weak, that is $g \ll \omega_s, \omega_p$. In this case
\begin{equation}\label{eq07}
    \pi_M \approx \frac{\vert p_z \vert g^2}{(\omega_s - \omega_p)^2 +
    g^2}
\end{equation}
by using the definitions in (\ref{eq05}), and the maximal
purification amplitude exhibits a Cauchy-Lorentz resonance for
$\omega_p = \omega_s$. The transition from the maximally mixed state
to a pure state is obtained by choosing $\vert p_z \vert = 1$. In
Fig.~\ref{fig1} we present the results of a numerical analysis of
$\pi_M$, showing the behavior of the resonance curve for increasing
values of the coupling constant $g$. As intuition could suggest, the
ability of inducing specific transitions in the system $S$ by means
of $P$ increases with the strength of the coupling between $S$ and
$P$.

{\it Purification in open system dynamics.---} The incoherent
control protocol can be effectively used to increase the purity of
the states of an open system, counteracting the decohering effects
of the environment. To show this, we adopt the previous model for
the free dynamics of the composite system $S + P$, and we add a
dissipative contribution:
\begin{equation}\label{eq08}
    \dot{\rho}_T (t) = -i [H_T, \rho_T (t)] + {\cal D} [\rho_T (t)],
\end{equation}
where we assume that the additional term is a Lindblad contribution
representing isotropic decoherence,
\begin{eqnarray}\label{eq09}
    {\cal D} [\rho_T (t)] &=& \gamma_s \Bigl( \sum_i \sigma_i^S \otimes {\mathbb
    I} \rho_T (t) \sigma_i^S \otimes {\mathbb I} - \rho_T (t) \Bigr) +
    \nonumber \\
    &+& \gamma_p \Bigl( \sum_i {\mathbb I} \otimes \sigma_i^P \rho_T (t)
    {\mathbb I} \otimes \sigma_i^P - \rho_T (t) \Bigr),
\end{eqnarray}
where $\gamma_s$, $\gamma_p$ are the decay rates for $S$ and $P$
respectively, and we neglect any environmental coupling between
system and probe.

\begin{figure}[t]
\begin{center} 
  \includegraphics[width=8cm]{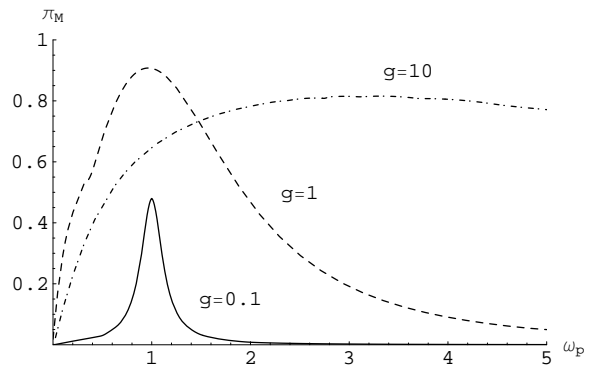} \\
 \caption{\footnotesize Maximal attainable purification $\pi_M$ for the maximally
 mixed state, with isotropic decoherence. In general $\pi_M$ is decreased by
 the environmental action, and it still exhibits a resonant behavior. We observe
 that the peak is about $\omega_s = \omega_p$ only when the interaction is weaker
 or comparable to the free contribution, otherwise it moves to $\omega_p > \omega_s$.
 We have assumed $\omega_s = 1$ and $\gamma_s = \gamma_p = 0.01$.}\label{fig2}
\end{center}
\end{figure}

The results of our numerical analysis are shown in Fig.~\ref{fig2}.
Also in this case the best choice of the probe is $\vert p_z \vert =
1$. The general effect of the interaction with the surrounding
environment is a decrease of $\pi_M$ depending on $g$. By comparing
Fig.~\ref{fig1} and Fig.~\ref{fig2}, we notice that this effect is
suppressed when $g$ is of the order of $\omega_s$ and $\omega_p$,
stronger otherwise. We observe that the peak of the resonance moves
from $\omega_p = \omega_s$ when the coupling constant $g$ is
dominant. Moreover, the dissipative dynamics produces a drastic
decrease of $\pi_M$ for $\omega_p \rightarrow 0$.

These effects can be understood by analyzing the different
contributions to the dynamics of $S$. We can identify three time
scales. Two of them characterize the non-Markovian dynamics of $S$,
depending on the interaction with $P$. From (\ref{eq06bis}), they
are given by $(\alpha + \beta)^{-1}$ (fast oscillations) and
$(\alpha - \beta)^{-1}$ (slow oscillations). Finally, there is the
characteristic time for the damping induced by the environment,
given by $\gamma_s^{-1}$.

In order to have an effective purification, the environment induced
relaxation should be negligible on a time scale of the order of the
slow oscillations, that is $\gamma_S^{-1} \gg (\alpha -
\beta)^{-1}$, otherwise the purification process is weakened by
dissipation. Differently from the closed system case, the
purification process is not improved by increasing the strength of
the coupling between $S$ and $P$, since in this case $\alpha
\thickapprox \beta \thickapprox g$ and then $(\alpha - \beta)^{-1}
\rightarrow + \infty$. In particular, this explains the decrease of
$\pi_M$ when $g$ is increased, in the open system case. In other
words, by increasing $g$ the purification time becomes large and the
dissipative effects come into play. Conversely, in the closed system
case the purification process takes advantage of a large $g$ since a
large purification time does not represent a problem.

A similar result holds for $\omega_p \rightarrow 0$ since $\alpha
\thickapprox \beta$ as well. This explains the fast decrease of
$\pi_M$ near $\omega_p = 0$. Note that the same pattern is expected
for $\omega_p \rightarrow + \infty$, assuming that $g \gg \omega_s$,
even if this is not apparent from Fig.~\ref{fig2}.

\begin{figure}[t]
\begin{center} 
  \includegraphics[width=8.5cm]{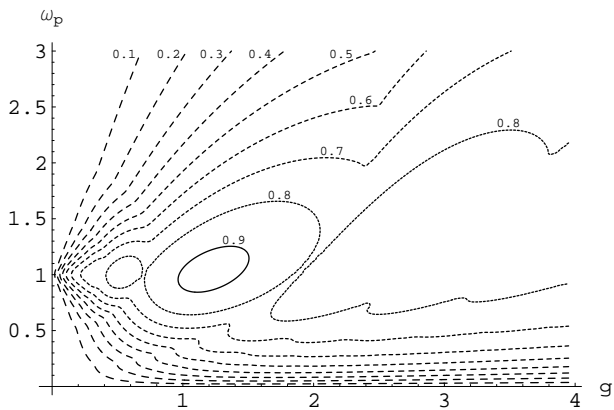} \\
 \caption{\footnotesize Contour plot of the maximal attainable purification
 $\pi_M$ in the $g-\omega_P$ plane, for the maximally mixed state, with
 isotropic decoherence. The dynamical parameters are fixed as in
 Fig.~\ref{fig2}. The global maximum is given by $\pi_M \thickapprox 0.92$,
 with $g \thickapprox 1.2$ and $\omega_p \thickapprox 1.0$.}\label{fig3}
\end{center}
\end{figure}

In Fig.~\ref{fig3} we present the contour plot of $\pi_M$ in the
$g-\omega_P$ plane for a particular choice of the dynamical
parameters. The surface levels suggest a non trivial relation among
$\pi_M$, $g$ and $\omega_p$, in particular there are several local
maxima.

{\it Conclusions.---} We have explored the purification process of a
quantum system $S$ based on its interaction with an auxiliary system
$P$, in both the closed and open system frameworks. We have
considered the case of two-level systems $S$ and $P$, and assumed
particular forms of the free dynamics, of the interaction between
$S$ and $P$, and of their interaction with the environment. We have
found that the purification process exhibits a resonant behavior
that depends on the energy difference between the system and the
probe. When the interaction between $S$ and $P$ is weak, this
resonance is described by a Lorentz-Cauchy curve. We have shown that
complete purifications are possible in the closed system case. In
the open system case, the attainable purity is reduced, and it
reaches a maximum for optimal values of the parameters
characterizing the dynamics.

The probe-mediated protocol differs from the existing purification
techniques, based on measurement or cooling, and it has a different
range of applicability. From this point of view, it complements the
existing methods. Usually, in measurement-based schemes, an indirect
measurement is needed to implement feedback control. In our case,
measurement does not enter the procedure (unless the initial state
of $P$ is prepared by a measurement), and there is not feedback.
Similarly to the cooling procedures, in our approach the aim is to
transfer the purity between different systems. However, this
transition has a completely different origin, since it depends on
the non-Markovian dynamics of $S$, not on a particular dissipative
mechanism. Therefore, the method is effective with any kind of
dissipation at work, and then it is of wider applicability.
Moreover, in our approach there is not coherent control on $S$.

The results obtained with the particular model described in this
work are not exceptional. In fact, we have found numerical evidence
of the resonance in the purification process with more general
Hamiltonian terms. Using $H_S = \vec{n} \cdot \vec{\sigma}^S$, $H_P
= \vec{m} \cdot \vec{\sigma}^P$, and varying the vectors $\vec{n}$
and $\vec{m}$, we have observed that the purification of the
maximally mixed state is optimal when $\vec{p}(u) = \vec{m}$.
Further analysis will explore these more general cases as well as
general environments. Complete purifications are expected also in
the open system case, when dissipative models different from
isotropic decoherence are considered.


\end{document}